# A Novel Dual Modality Sensor With Sensitivities to Permittivity, Conductivity, and Permeability

Jorge R. Salas Avila, Kin Yau How, Mingyang Lu, and Wuliang Yin, *Senior Member, IEEE*

*Abstract*—In this paper, an electromagnetic sensor which can operate simultaneously in capacitive and inductive modalities with sensitivities to permittivity, conductivity, and permeability is developed, and a novel measurement strategy is proposed accordingly. The sensor is composed of two planar spiral coils with a track width of 4 mm, which promotes its capacitive mode. The capacitive coupling is measured in common mode, while the inductive coupling is measured in differential mode. In capacitive mode, the sensor is sensitive to changes in permittivity, i.e., the dielectric material distribution; while in inductive mode, it is sensitive to magnetically permeable material and electrically conductive material. Furthermore, it is demonstrated that the sensor can simultaneously measure dielectric and conductive materials. This novel sensing element has been designed and implemented. Experimental results verified its effectiveness in dual modality measurement.

*Index Terms*—Planar sensors, EM sensor, dual modality, eddy-current testing, combined sensing.

## I. INTRODUCTION

EVALUATION of materials by using electric or magnetic fields has been extensively performed for various inspection purposes, such as failure detection, quality assurance and material composition inspection [1], [2]. The selection of the measurement method is determined by the fundamental electrical and magnetic properties of the material of interest i.e. permittivity, conductivity and permeability [3]. Capacitance measurements are appropriate for evaluating dielectric materials; for example, planar capacitance sensors have been used to inspect variations in dielectric properties of materials [4], [5]. Magnetic induction / eddy-current testing is suitable for evaluating and inspecting conductive/permeable materials with many different coil configurations having been investigated, including planar spiral coils [6].

Measuring the change in both capacitance and mutual inductance with a suitable sensor gives the possibility of inspecting the fundamental electrical and magnetic properties (permittivity, conductivity and permeability) with a single sensor. Therefore, insulators, conductors and composite materials can be inspected with one sensor.

Manuscript received September 19, 2017; accepted October 23, 2017. Date of publication October 27, 2017; date of current version December 7, 2017. The work of J. R. Salas Avila was supported by the National Council of Science and Technology (CONACYT) of Mexico. The associate editor coordinating the review of this paper and approving it for publication was Dr. Patrick Ruther. *(Corresponding authors: Jorge R. Salas Avila; Wuliang Yin.)*

The authors are with the School of Electrical and Electronic Engineering, University of Manchester, Manchester M13 9PL, U.K. (e-mail: jorge.salasavila@postgrad.manchester.ac.uk; kin.how@student.manchester.ac.uk; mingyang.lu@postgrad.manchester.ac.uk; wuliang.yin@manchester.ac.uk).

Attempts of combining capacitive and inductive measurements have been reported previously. In [7] and [8], by switching between modes of operation or multiplexing, the presence of conductive and dielectric materials is detected with a dual mode sensor, but the sensor is still a physical combination of two sensors (separate capacitive and inductive elements). A printed sensor was reported in [9]; by identifying the predominant sensor response above and below the resonant frequency, it was possible to distinguish between conductive and dielectric materials. In [10], meander and mesh planar sensors were employed for inspection of conductive and dielectric materials; the effects of some dielectric samples on the transfer impedance using frequencies up to hundreds of megahertz were reported. This sensor is sensitive to both conductive and dielectric materials, but was not capable of determining both properties simultaneously.

In this paper, we present a novel sensor which inherently is a dual inductive/capacitive sensing element and thus is sensitive to changes in conductivity, permittivity and permeability; and importantly, inductive/capacitive effects can be separated by using different modes of measurement. In differential mode, the change in mutual inductance is measured; and in common mode, the change in capacitive coupling is measured. Measurements can be taken in differential and common modes simultaneously with an impedance analyser with a suitable configuration, therefore, the sensor can work simultaneously in inductive and capacitive modes and there is no need for switching between different sensing elements. This allows fast measurements to be carried out and avoids the associated disadvantage in [7], i.e. the settling time and system stability need to be considered associated with the switching and a programmable delay has to be introduced to avoid interference. Moreover, the designed planar sensor has some advantages including, easy manufacturing, good repeatability, low cost as in [11], and can be built of flexible materials for inspection of irregular surfaces as in [6].

The sensor was designed and built, and experimental results for measuring conductive, dielectric and permeable materials are presented. We tested the sensor with a range of materials and combinations: such as water, air, plastic plates, copper plates and ferrite rings. In addition, the sensor was coated and results are presented for an immersion experiment.

## II. SENSOR DESIGN

The sensor is composed of two planar spiral coils printed on a PCB. Fig. 1 depicts the layout of the sensor. The trace width is 4 mm, and the gap between the traces is 1 mm;

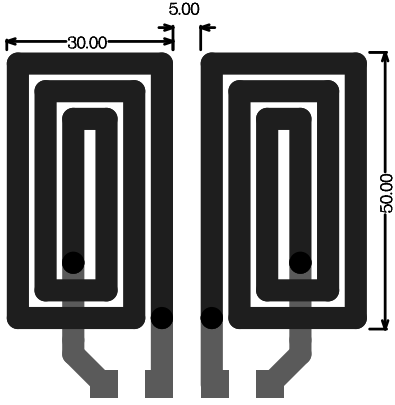

Fig. 1. Sensor layout. Units are in millimetres.

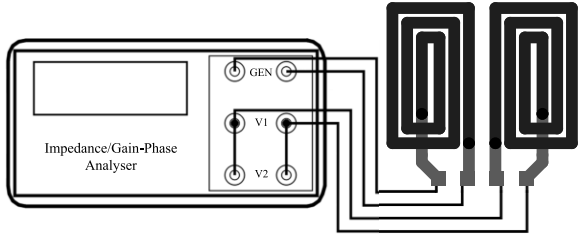

Fig. 2. Connections between the sensor and the impedance analyser.

the separation between the nearest traces of the coil pair is 5 mm, and the distance between the centres of the coil pair is 41 mm. The PCB substrate is made of 1.6 mm FR-4, with a relative permittivity value of 4.4.

The self-inductance of each coil is ∼320 nH, mutual inductance 20 nH and direct coupling capacitance 1.56 pF at 100 kHz measured with an impedance analyser (SL 1260). The same instrument was used in the following experiments; connections between the sensor and the impedance analyser are shown in Fig. 2. The instrument has two measurement channels which can be independently configured as differential and common modes; thus, differential and common mode measurements can be taken simultaneously with this configuration.

## III. SENSING MODES AND MEASUREMENTS MODES

### A. Inductive Sensing Mode

Currents flowing in the tracks on the excitation side produce magnetic field, which induces voltage in the receiving side due to magnetic induction. So, the same planar structure that is used for capacitive measurements can be treated as coils for magnetic induction measurement.

The conductivity and magnetic permeability of the sample affect the magnetic induction due to eddy currents and magnetic polarisation, and the effects can be measured through the induced voltage across the receiver coil [12]. As the magnetic field depends on the coil geometry, the sensitivity of the sensor is intrinsically related to its geometry. The analytical solution for the change in impedance of a planar circular spiral coil can be derived from Dodd and Deeds theory as presented by Ditchburn [6]. Circular and rectangular geometries for planar coils have been compared due to its similar behaviour [13].

The Dodd and Deeds analytical solution describes the inductance change of an air-core coil pair caused by a metallic plate for both non-magnetic and magnetic cases. The difference in the complex mutual inductance is $\Delta L(\omega) = L(\omega) - L_A(\omega)$ where the coil inductance above a plate is $L(\omega)$, and $L_A(\omega)$ is the inductance in free space.

In the region between l1 and l2, the vector potential can be expressed as (1), where N1, N2 denote the number of turns in the excitation and pickup coil; $\alpha$ is a spatial frequency variable; $\mu 0$ denotes the permeability of free space; le1 and le2 denote the height of bottom and top of the excitation coil; while lp1 and lp2 denote the height of bottom and top of the pickup coil; re1 and re2 denote the inner and outer radii of the excitation coil; while rp1 and rp2 denote the inner and outer radii of the pickup coil; and c denotes the thickness of the plate in Fig. 3. J(x) is a first-order Bessel function of the first kind. I(x1, x2) represent the production of J(x) for radii of x1 and x2.

$$
\begin{aligned}
\mathbf{A}^{(1,2)}&(r, z) \\
&= \frac{\mu_0 \mathrm{IN1}}{(r_{e_2} - r_{e_1})(l_{e_2} - l_{e_1})} \int_0^\infty \frac{1}{\alpha^3} \mathbf{I}(r_{e_2}, r_{e_1}) \cdot \mathbf{J}(\alpha r) \\
&\times [2 - e^{\alpha(z - l_{e_2})} - e^{\alpha(z - l_{e_1})}] \\
&+ e^{-\alpha z}\left(e^{-\alpha l_{e_1}} - e^{-\alpha l_{e_2}}\right) \\
&\frac{(\alpha_1 + \mu\alpha)(\alpha_1 - \mu\alpha) - (\alpha_1 + \mu\alpha)(\alpha_1 - \mu\alpha)e^{2\alpha_{1}c}}{-(\alpha_1 - \mu\alpha)(\alpha_1 - \mu\alpha) + (\alpha_1 + \mu\alpha)(\alpha_1 + \mu\alpha)e^{2\alpha_{1}c}}] \mathrm{d}\alpha
\end{aligned}
\tag{1}
$$

The voltage induced in the reciver with a single turn can be expressed as

$$
V = j\omega \int_S \mathbf{A}(r, z) \, ds = j\omega \int_S \mathbf{A}(r, z) \, r_p \cos(\varphi) \mathrm{d}\theta
\tag{2}
$$

where $\varphi = \theta + \mathrm{tg}^{-1}(r_p \sin\theta / (w - r_p \cos\theta))$ is the angle between A and ds; s denotes the transect area in the pickup coil; while $r = \sqrt{r_p^2 \sin^2\theta + (w - r_p\cos\theta)^2}$ is the distance between O and ds.

Considering (1) and (2), the induced voltage on the receiver can be yielded in (3).

$$
\begin{aligned}
V&{}={} \frac{\mathrm{N1N2j}\omega\mu_0 \mathrm{I}}{(r_{e_2} - r_{e_1})(l_{e_2} - l_{e_1})(r_{p_2} - r_{p_1})(l_{p_2} - l_{p_1})} \\
&\times \int_0^\infty \int_0^{2\pi} \int_{r_{p_1}}^{r_{p_2}} \cos\left(\theta + \mathrm{tg}^{-1}\left(\frac{r_p\sin\theta}{w - r_p\cos\theta}\right)\right) \frac{1}{\alpha^3} \mathbf{I}(r_{e_2}, r_{e_1}) \\
&\mathbf{J}(\alpha\sqrt{r_p^2\sin^2\theta + (w - r_p\cos\theta)^2})\{2(l_{e_2} - l_{e_1}) \\
&- \frac{1}{\alpha}[2e^{-\alpha(l_{e_2} - l_{e_1})} - 2 + \left(e^{-\alpha l_{e_1}} - e^{-\alpha l_{e_2}}\right)^2 \\
&\frac{(\alpha_1 + \mu\alpha)(\alpha_1 - \mu\alpha) - (\alpha_1 + \mu\alpha)(\alpha_1 - \mu\alpha)e^{2\alpha_{1}c}}{-(\alpha_1 - \mu\alpha)(\alpha_1 - \mu\alpha) + (\alpha_1 + \mu\alpha)(\alpha_1 + \mu\alpha)e^{2\alpha_{1}c}}\}]\mathrm{d}r_p\mathrm{d}\theta\mathrm{d}\alpha
\end{aligned}
\tag{3}
$$

Consequently, the mutual inductance between the air-cored coil pair can be presented by dividing the induced voltage by the current flowing through the excitation coil, as shown in (4).

$$
\begin{aligned}
L &= \frac{N1N2\mu_0}{(r_{e_2} - r_{e_1})(l_{e_2} - l_{e_1})(r_{p_2} - r_{p_1})(l_{p_2} - l_{p_1})} \\
&\times \int_0^\infty \int_0^{2\pi} \int_{r_{p_1}}^{r_{p_2}} \cos\left(\theta + \mathrm{tg}^{-1}\left(\frac{r_p \sin\theta}{w - r_p \cos\theta}\right)\right) \frac{1}{\alpha^3} \mathbf{I}(r_{e_2}, r_{e_1}) \\
&\quad \mathbf{J}(\alpha\sqrt{r_p^2 \sin^2\theta + (w - r_p\cos\theta)^2})\{2(l_{e_2} - l_{e_1}) \\
&\quad - \frac{1}{\alpha}[2e^{-\alpha(l_{e_2} - l_{e_1})} - 2 + \left(e^{-\alpha l_{e_1}} - e^{-\alpha l_{e_2}}\right)^2 \\
&\quad \frac{(\alpha_1 + \mu\alpha)(\alpha_1 - \mu\alpha) - (\alpha_1 + \mu\alpha)(\alpha_1 - \mu\alpha)e^{2\alpha_1 c}}{-(\alpha_1 - \mu\alpha)(\alpha_1 - \mu\alpha) + (\alpha_1 + \mu\alpha)(\alpha_1 + \mu\alpha)e^{2\alpha_1 c}}]\}dr_p d\theta d\alpha
\end{aligned}
$$

$$(4)$$

### B. Capacitive Sensing Mode

With large surface track width, each of the planar coils acts as a capacitive plate, where one is the transmitter and other the receiver. Therefore the capacitive sensing mechanism is similar to that of a two coplanar plate configuration [14]. When a sinusoidal voltage is applied to the transmitter, a potential difference is established and thus a capacitive coupling developed. Introducing permittivity change in the sensing area will perturb the established potential distribution and hence the capacitive coupling, the change of which can then be measured.

Depending on the nature of the sample, different effects are expected as discussed in [15]. A grounded object reduces the electric flux reaching the receiver due to a leakage through the newly formed ground path, and therefore will reduce the capacitive coupling; a floating sample with a higher permittivity generally increases the capacitive coupling. These effects are referred as shunt mode and transmission mode respectively [14]. Both effects were observed in our sensor, but the interest of this work is for the latter case where the sample is electrically floating.

Goss *et al.* [16] identified six coupling mechanisms for an excitation/detection coil pair with a sample in-between. It was stated that the potential difference between the coils, the surface area of the target, and the direct capacitive coupling between the coils strongly influence the capacitive excitation – capacitive detection mode. While in magnetic inductive measurements, the capacitive coupling effect needs to be minimised, the sensor developed here intentionally exploits this effect. By using a large track width, a significant direct capacitive coupling between the tracks develops. The track width is limited by the overall sensor size and therefore a track width of 4 mm was selected.

2-D finite-element simulations were carried out to explore the sensitivity distribution of the sensor over a sample. Both coils, excitation and detection, were segmented in 6 traces $T_A$ to $T_F$ and $D_A$ to $D_F$ as shown in Fig. 4. Treated as a coil, a sinusoidal potential is applied to the excitation element. For simulation, different potentials were assigned to the excitation coil traces due to the fact that it is the only load for the signal generator and the voltage drop must occur along the excitation

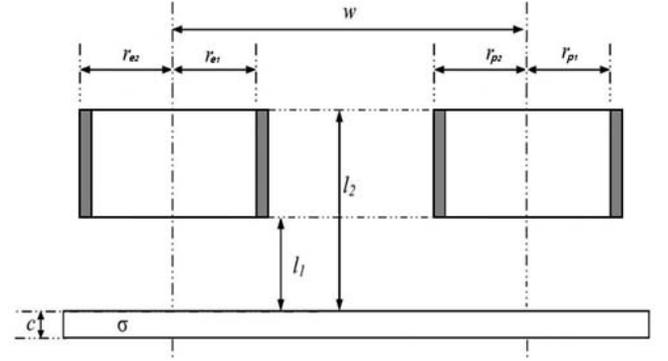

Fig. 3. Dodd and Deeds simulations setup.

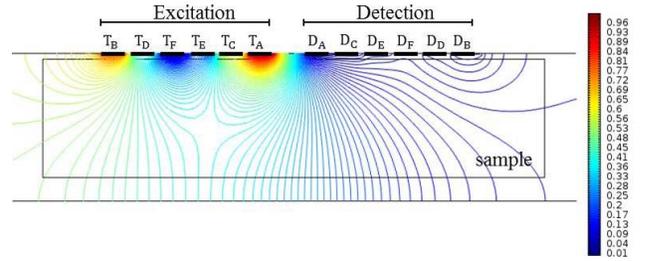

Fig. 4. Electric field distribution. Units are in volts.

tracks. The excitation coil potentials going from 1 V to 0 V. Assuming no inductive coupling, as in conventional capacitive measurements [17], all the traces of the detection coil are at the same potential. For the sensor working simultaneously in inductive and capacitive modes, a potential difference exists over the detection coil between its different segments due to inductive coupling. Different potentials were assigned to the detection coil; considering the sensor inductive coupling coefficient, the receiver coil potentials were set to 1/16 of the excitation coil potentials. The electric field distribution for an inductively coupled receiver is shown in Fig. 4. The relative permittivity of the sample was set to one. It is worth noting that the capacitive coupling between segments is independent of the potential set up. From the previous simulation, it can be observed that individual capacitive coupling between each segment of the transmitter and the receiver develops; i.e. between segment $T_A$ and the receiver segments $D_A$ to $D_F$, $T_B$ and the receiver segments $D_A$ to $D_F$, and so on; Table I shows the simulation values for the segments $T_A$, $T_B$ and $T_C$. Simulation results of the capacitance between the individual transmitter tracks $T_A$, $T_B$ and $T_C$ with the receiver indicate that the nearest track $T_A$ has the strongest coupling with the receiver. Table II shows the developed capacitance between the nearest track of the excitation coil $T_A$ and the segments $D_A$, $D_B$ and $D_C$. Therefore, the strongest capacitive coupling is between the adjacent tracks of the excitation and detection coils i.e. between the segments $T_A$ and $D_A$. The sensor sensitivity distribution to permittivity calculated according to the E dot E formulation [18], [19] is shown in Fig. 5. The centre of the sensor is positioned at the coordinate x = 0 mm, y is the distance between the sensor and the sample. As expected

CAPACITANCE BETWEEN A TRANSMITTER TRACK AND THE RECEIVER

| Coupling | Relative permittivity of the sample | | |
|---|---|---|---|
| | 1 | 3 | 80 |
| $T_A$-Receiver | 446 fF | 622 fF | 1201 fF |
| $T_B$-Receiver | 160 fF | 215 fF | 927 fF |
| $T_C$-Receiver | 159 fF | 278 fF | 955 fF |

TABLE II
CAPACITANCE BETWEEN INDIVIDUAL TRACKS

| Coupling | Relative permittivity of the sample | | |
|---|---|---|---|
| | 1 | 3 | 80 |
| $T_A$-$D_A$ | 266 fF | 316 fF | 291 fF |
| $T_A$-$D_B$ | 33 fF | 49 fF | 185 fF |
| $T_A$-$D_C$ | 66 fF | 109 fF | 205 fF |

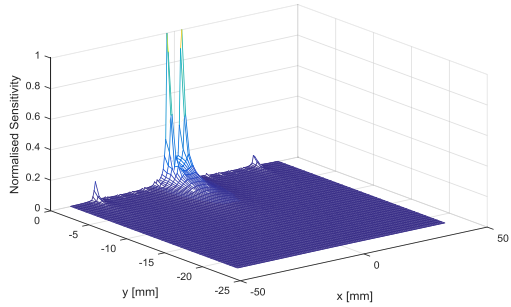

Fig. 5. Normalized sensitivity distribution of the sensor for capacitive sensing mode.

the sensitivity is concentrated in the centre region of the sensor.

Overall, the average coupling effect from all the segments is measured. Simulation results give an overall capacitance between the transmitter and the receiver of 1.34 pF which is in accordance with measurement results.

### C. Measurements Modes: Differential Mode, Common Mode, and Simultaneous Mode

An equivalent circuit of the sensor and a sample in-between is shown in Fig. 6 [16]. Coupling is both capacitive and inductive between the coils (direct coupling: Cd and Md) and through the sample (indirect coupling: Cs1, Ms1, Cs2 and Ms2). The track resistance and parasitic capacitance of the coils are not shown. Each of the planar sensors is treated as a coil, represented as L1 for the transmitter and L2 for the receiver. The target is modelled as an equivalent RLC parallel circuit, where R3 represents the losses due to eddy currents for a conductive sample; L3 is the inductive element related to the eddy currents; and C3 is the capacitive coupling element related to the displacement currents.

The change in mutual inductance between the transmitter and the receiver can be detected by measuring the differential voltage change at the two terminals of the receiver coil [12]. As stated by Equation (5), due to magnetic induction, the change in voltage $\Delta V$ is proportional to the changes in mutual inductance $\Delta M$, the current in the transmitter coil $I$ and the angular frequency $\omega$; $j$ is the imaginary unit.

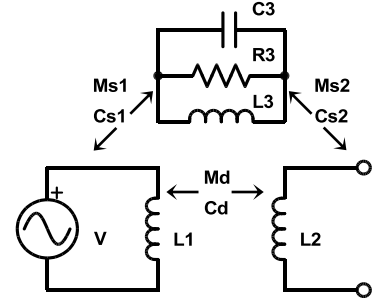

Fig. 6. Equivalent circuit of the sensor and a sample.

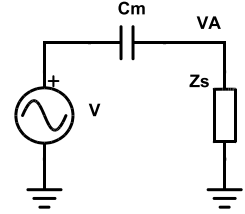

Fig. 7. Simplified common mode setup.

$$\Delta V = j\omega \Delta M I \tag{5}$$

The measured capacitance Cm is Cd in parallel with the series equivalent of Cs1, C3 and Cs2.

In order to separate the inductive and capacitive coupling effects, different measurement modes were used, i.e. common mode and differential mode. With common mode, the measurement is sensitive to the potential difference between the transmitter and receiver and therefore it is related to capacitive coupling. With differential mode, it is sensitive to the voltage difference between the receiver coil terminals and therefore related to inductive coupling.

In differential mode, the circuit can be treated as two mutually coupled coils.

In common mode, the circuit can be simplified as shown in Fig. 7. $V_A$ is the common mode voltage that is determined by the unknown capacitance Cm and the input impedance of the impedance analyser Zs. Zs can be treated as a constant RC parallel circuit once the measurement setup is fixed.

Simultaneous mode (simultaneous capacitive and inductive measurement) was developed in order for the proposed sensor to be able to sense conductivity, permittivity and permeability simultaneously. Common mode voltage sees the receiver coil as one conductive surface at a reference potential level at the point of connection. Therefore, paths for the movement of charge due to the potential difference between the transmitter and the receiver are created. However, common mode measurements do not force a uniform potential level on the receiver; hence a differential voltage across the receiver terminals due to inductive coupling can be measured simultaneously.

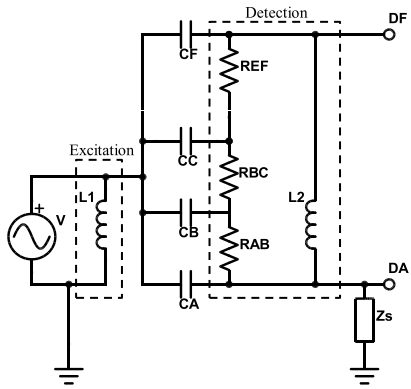

Fig. 8. Equivalent circuit of the sensor in simultaneous mode.

An equivalent circuit for simultaneous mode is shown in Fig. 8. L1 and L2 represent the excitation and receiver elements respectively. The potential difference between points DA and DF at the receiver due to inductive coupling is represented with resistors RAB to REF and coil L2. Capacitances CA to CF represent the capacitive coupling at different points on the receiver track. Zs is the input impedance of the impedance analyser.

Differential mode voltage is the voltage difference between the points DA and DF. As shown above, the strongest capacitive coupling effect occurs between the nearest tracks of the transmitter and the receiver. Therefore, when taking common mode measurement, the point at which the voltage is measured is at DA.

Considering the equivalent circuit of Fig. 8, differential mode measurement $\Delta V = V_{DF} - V_{DA}$ contains inductive coupling. Common mode voltage can be defined as $V_A = I_C Z_S$, where $I_C$ is the current from the transmitter to the receiver due to overall capacitive coupling i.e. the sum of all the currents $I_A$ to $I_F$ in the form of (6).

$$I_A = j\omega \left( C_{TA-DA} V_{TA-DA} + C_{TB-DA} V_{TB-DA} \right.$$
$$\left. + C_{TC-DA} V_{TC-DA} + \ldots \right) \quad (6)$$

## IV. EXPERIMENTAL SETUP AND SAMPLES

An impedance/gain-phase analyser SL 1260 was used to carry out the measurements. The instrument has a signal generator output and two input voltage measurement channels that can be configured as either common mode or differential mode. As shown in Fig. 2, the excitation coil was connected to the signal generator and the receiver coil terminals were connected to both voltage channels in parallel; channel one was configured as differential mode and channel two as common mode. Fig. 9 shows the schematic connections between the equivalent circuit and the instrument for each channel separately. Fig. 9(a) represents common mode and Fig. 9(b) differential mode. R1 and R2 in both figures represent the input impedance of the impedance analyser with the values taken from the instrument manual.

As can be seen from Fig. 9(a), when common mode is selected, only one of the terminals is internally connected for measuring purposes. Thus, the receiver coil is represented as only one plate. Cm represents the coupling capacitance between the sensor pair. From Fig. 9(b) it can be seen that the measured voltage in differential mode corresponds to the differential voltage between the two terminals of the receiver coil.

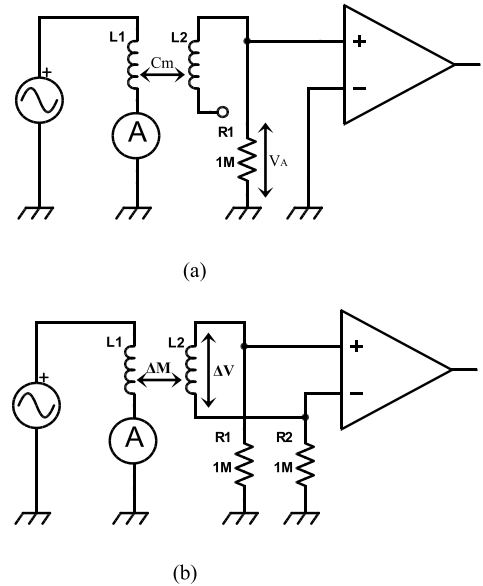

Fig. 9. Instrument connections for each mode separately. (a) Common mode. (b) Differential mode.

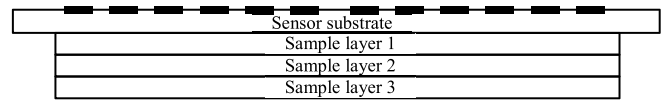

Fig. 10. Setup for measuring sample layers.

Once all connections are made, simultaneous data of both channels (one in differential mode and the other in common mode) can be obtained with the instrument.

Samples with different electromagnetic properties were prepared to test the behaviour of the sensor. In order to test the inductive coupling, a set of conductive samples with different thickness was created by stacking 1-5 copper foil layers. Each layer has a thickness of 60 $\mu$m. A set of high permeable samples (nickel-zinc 10 mm x 8 mm ferrite rings) was used to introducing permeability changes. Ferrite rings were centred between the transmitter and the receiver. Lift-off is 1.6 mm.

For testing the capacitive effect, plastic sheets were measured and a water immersion experiment was carried out. Lift-off for plastic sheets is 1.6 mm; each plastic plate is 1.5 mm thickness and has dimensions of 30 mm by 65 mm. For the immersion experiment, the sensor was coated; the surface area of the water volume was constant, and the height is linearly related to the volume, approximately 1 mm for each 10 mL of water once the sensor is fully submerged. For the following discussion, relative permittivity values of 1 for air, 3 for plastic samples and 80 for water samples are assumed; which are typical values for these materials. A representation of the samples positioning is shown in Fig. 10. Excitation frequency was set to 1 MHz for common mode and simultaneous experiments; and to 100 kHz for the differential mode experiments.

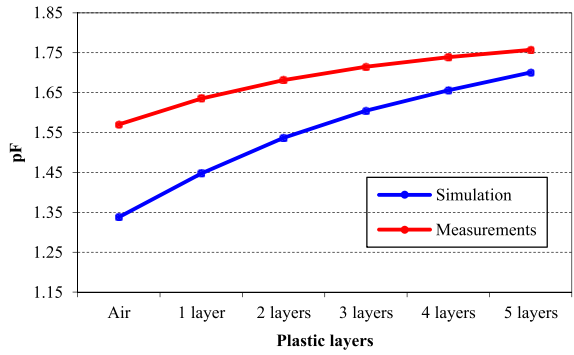

Fig. 11. Capacitance in common mode: plastic samples.

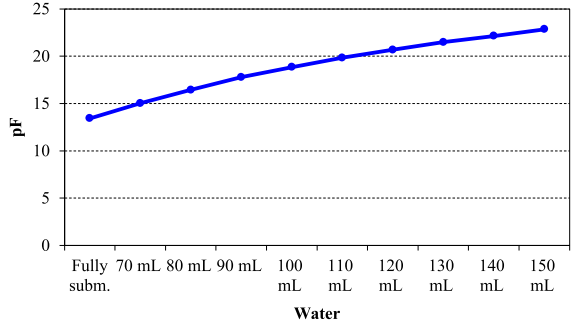

Fig. 12. Capacitance in common mode: water immersion. Fully submerged corresponds to 60 mL.

## V. RESULTS

Results were obtained first for capacitive sensing/common mode measurement, then for inductive sensing/differential mode, and lastly for the simultaneous mode.

A plot for capacitive-sensing/common-mode-measurement is shown in Fig. 11. The first datum, labelled Air, is the inter-plates capacitance value of the sensor in air; in this case, the capacitive coupling is through the PCB substrate and air. As the thickness of the plastic plate increases, the measured capacitance increases as expected.

Results of a 2D finite element simulation are also shown in Fig. 11. The capacitive coupling increases as the plastic sample thickness increases, same trend as the in the experiments. The capacitance range is 1.34 pF to 1.7 pF for simulation and 1.57 pF to 1.76 pF for experiments. The error can be attributed to the 2D nature of the simulation.

Measurements for different volumes of water were also carried out in common mode. As shown in Fig. 12, the capacitive coupling increases with increasing volume of water as expected. A larger change in the measured capacitance due to the presence of water than plastic plates can also be seen, which is attributed to a much higher permittivity of water than that of the plastic sample.

To test inductive coupling in differential mode, copper samples with different thickness positioned at 5 mm away were measured at 100 kHz (Fig. 13). A reduction in the measured voltage is observed when the thickness of the sample increases. This reduction is in accordance with simulations using the method in Section III and the magnetic induction effect for highly conductive, nonmagnetic samples [12].

To test the simultaneous mode, two experiments were carried out:

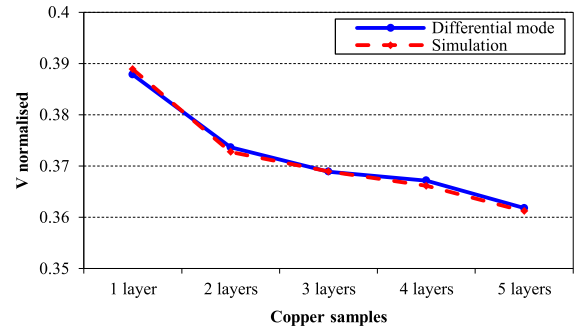

Fig. 13. Voltage change in differential mode: copper samples 5 mm away (100 kHz). $V_{normalized} = |V_{sample} - V_{air}|/V_{air}$.

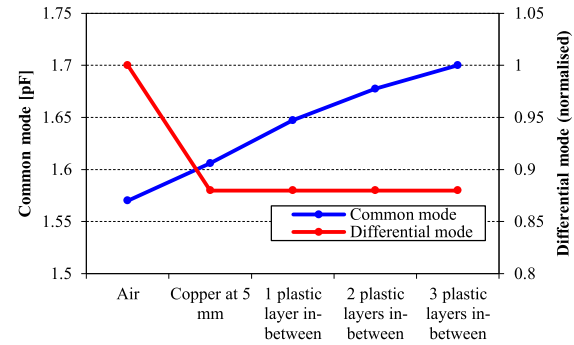

Fig. 14. Common mode and differential mode: plastic samples in-between a five layers copper sample positioned 5 mm away. $V_{normalized} = 1 + (V_{sample} - V_{air})/V_{air}$.

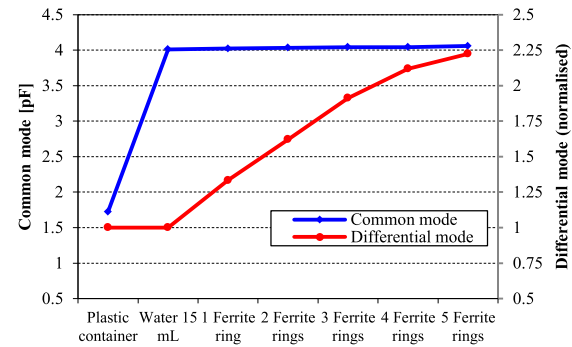

Fig. 15. Common mode and differential mode: water and ferrite rings. $V_{normalized} = 1 + (V_{sample} - V_{air})/V_{air}$.

1) A 300 $\mu$m copper plate sample was positioned 5 mm away from the sensor, and then, plastic samples of different thicknesses were placed in-between. As expected, the differential measurement remained the same but the common mode measurements increase with the thickness of the plastic plates (Fig. 14). Therefore, it was verified that the sensor can simultaneously operate in both modes.

2) Ferrite rings were introduced in a container with 15 mL of water (Fig. 15). The experiment demonstrates that differential mode measurement is sensitive to permeable materials (ferrite rings) but not to materials with permittivity (water); common mode measurement is mainly sensitive to materials with permittivity (water) with a change of capacitance of 2.3 pF, but only weakly sensitive to permeability (ferrite rings) with a small change of 0.05 pF.

## VI. Conclusions and Future Work

In this work, a novel dual modality sensor and the corresponding measuring strategy are presented. The sensor acts as a spiral coil pair, and as a planar capacitive sensor pair depending on the measurement mode. Equivalent circuits of the sensor were utilised to analyse the sensor response and develop the measurement strategy for each mode. Results from measurements indicate that in differential mode, the change in mutual inductance is measured; in common mode, the change in capacitance coupling is measured; and that simultaneous measurements for inductive and capacitive coupling can be performed. Tests also suggest that the sensor is sensitive to conductivity and permeability in differential mode and permittivity in common mode. Therefore, this sensor and the measurement strategy have the potential to inspect insulators, conductors and composite materials. The versatility of the sensor is also demonstrated with an immersion experiment. Future research will focus on optimisation of sensor geometries for specific measurement applications such as inspection of composite materials in NDT [20], [21] and multiphase flow measurements; where components with conductivity, permittivity and permeability are present. In addition, a custom instrument based on FPGA [22]–[24] will be built to replace the commercial instrument (SL 1260) to implement the measurement strategy.

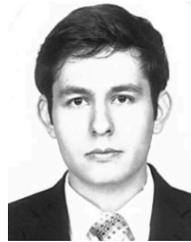

**Jorge R. Salas Avila** received the B.S. degree in electronic and computer engineering from the Monterrey Institute of Technology and Higher Education, Chihuahua, Mexico, in 2010, and the M.Sc. degree in informatics technology from the Monterrey Institute of Technology and Higher Education, Monterrey, Mexico, in 2013. He is currently pursuing the Ph.D. degree in electrical and electronic engineering with the University of Manchester, Manchester, U.K. His current research interests include instrumentation, electromagnetic sensors, FPGA-based digital instruments and non-destructive testing.

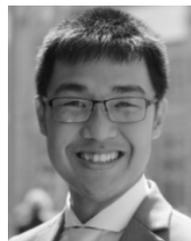

**Kin Yau How** is working towards the M.Eng. degree in electronic engineering at the School of Electrical and Electronic Engineering, University of Manchester, U.K., Manchester. His main area of research interest in sensors design and devising architectures for real-time magnetic imaging instruments.

**Mingyang Lu** is working towards the Ph.D. degree at the School of Electrical and Electronic Engineering, University of Manchester, mainly working on developing a FEM model to solve EM simulation taking into account random geometry and material properties (including microstructure) under the supervision of Wuliang Yin. His research interest is in developing software to increase efficiency of simulations to avoid remeshing, for example, to consider a moving sensor as a moving effective field above a flaw (non-destructive testing application).

**Wuliang Yin** (M'05–SM'06) was appointed as a MT (www.MT.com) Sponsored Lecturer with the School of Electrical and Electronic Engineering, University of Manchester, in 2012, and was promoted to Senior Lecturer in 2016. He has published one book, more than 180 papers and was granted more than ten patents in the area of EM sensing and instrumentation. He is a recipient of the 2014 and 2015 Williams Award from the Institute of Materials, Minerals and Mining for his contribution in applying EM imaging in the steel industry. He also received a Science and Technology Award from the Chinese Ministry of Education in 2000.